\RequirePackage{fix-cm}
\documentclass[smallextended]{svjour3}       
\smartqed  
\usepackage{graphicx}
\begin{document}

\title{Influence of periodic orbits on the formation of giant planetary systems
}
%


\titlerunning{Periodic orbits and formation of giant planetary systems}        

\author{Anne-Sophie Libert, Sotiris Sotiriadis and Kyriaki I. Antoniadou         
}


\institute{A.-S. Libert \and S. Sotiriadis \and K. I. Antoniadou \at
              naXys - Department of Mathematics, University of Namur, 8 Rempart de la Vierge,\\ 5000 Namur, Belgium  \\
              \email{anne-sophie.libert@unamur.be, sotiris.sotiriadis@unamur.be, \linebreak kyriaki.antoniadou@unamur.be}           
}

\date{Received: date / Accepted: date}

\maketitle

\begin{abstract}

The late-stage formation of giant planetary systems is rich in interesting dynamical mechanisms. Previous simulations of three giant planets initially on quasi-circular and quasi-coplanar orbits in the gas disc have shown that highly mutually inclined configurations can be formed, despite the strong eccentricity and inclination damping exerted by the disc. Much attention has been directed to inclination-type resonance, asking for large eccentricities to be acquired during the migration of the planets. Here we show that inclination excitation is also present at small to moderate eccentricities in two-planet systems that have previously experienced an ejection or a merging and are close to resonant commensurabilities at the end of the gas phase. We perform a dynamical analysis of these planetary systems, guided by the computation of planar families of periodic orbits and the bifurcation of families of spatial periodic orbits. {We show that inclination excitation at small to moderate eccentricities can be produced by (temporary) capture in inclination-type resonance and the possible proximity of the non-coplanar systems to spatial periodic orbits contributes to maintaining their mutual inclination over long periods of time.} 

\end{abstract}

\section{Introduction}
\label{intro}
To explain the diversity in eccentricity and inclination of the exoplanets, planet-planet interactions during migration in the protoplanetary disc are commonly invoked (\cite{Moorhead2005}, \cite{Moeckel2008}, \linebreak \cite{Matsumura2010}, \cite{Libert2011a}, \cite{Moeckel2012}). Two main mechanisms producing inclination increase have been identified in this scenario: planet-planet scattering and inclination-type resonance. Regarding the latter, during the disc-induced orbital migration of the planets, mean-motion resonance (MMR) capture takes place. As the planets continue to migrate while in MMR, their eccentricities increase, and when their values become high enough, the system can enter an inclination-type resonance (the resonant angle is a combination of the mean longitudes and the longitudes of the ascending node), which induces rapid growth of the inclinations (e.g. \cite{Thommes2003}, \cite{Libert2009}, \cite{Libert2011b},  \cite{Teyssandier2014}). 

Inclination-type resonance has first been observed by \linebreak \cite{Thommes2003} for the 2/1 MMR. \cite{Libert2009} have shown that capture into other MMRs (e.g. 3/1, 4/1 and 5/1) can also lead to inclination excitation when eccentricity damping is not very strong, in order for the eccentricity of one planet to exceed $\sim 0.4$. This empirical observation has been analytically confirmed for elliptic orbits for the 2/1 and 3/1 MMRs by \cite{Voyatzis2014}. 
This work has shown that inclination-type resonance is associated with the existence of vertical critical orbits along the planar family of resonant periodic orbits, where families of spatial periodic orbits bifurcate. Inclination-type resonance has also been observed for three-body resonances (e.g. the Laplace resonance) in \cite{Libert2011b}. Let us note that no inclination damping was considered in the above-mentioned works; the influence of inclination damping on the previous results has been studied in \cite{Teyssandier2014} and \cite{Sotiriadis2017}.

Recently, \cite{Sotiriadis2017} have performed extensive $n$-body simulations of three giant planets in the late stage of the disc, taking into account Type-II migration, the damping of planetary eccentricity and inclination (fitted from the hydrodynamical simulations of \cite{Bitsch2013}), and an exponential decrease of the disc mass. Starting from quasi-circular and quasi-coplanar orbits, their simulations show a very good agreement with the semi-major axis and eccentricity distributions of the detected giant planets. Despite the strong eccentricity and inclination damping induced by the gas disc, a significant proportion of highly mutually inclined systems are formed {($\sim5\%$ of the systems have a pair of planets with mutual inclination higher than $10^{\circ}$)}. While the majority of the three-dimensional (3D) systems formed in \cite{Sotiriadis2017} result from planet-planet scattering or orbital instability, $30\%$ of the 3D systems have final configurations influenced by mean-motion resonances. Half of them are captured in a three-body inclination-type resonance during the migration. In the remaining half of the systems, a planet-planet scattering event takes place, leading to the ejection/merging of one of the planets, and the final configuration of the two-planet system is found close to a resonant commensurability. In the following, we will refer to these systems as IRTP systems (for Inclined and Resonant Two-Planet systems). In the present work, we aim to analyse the dynamical evolution of the IRTP systems, and reveal the inclination-growth mechanisms that produce the mutual inclinations of their orbits. Moreover, we will show that the spatial periodic orbits have a significant influence on the final architectures of the systems.     

In Section \ref{sec:2}, we describe the orbital parameters of the IRTP systems formed in \cite{Sotiriadis2017}. Three evolutions showing inclination increase are presented in Section \ref{sec:3}, and the dynamical mechanism producing inclination excitation at small to moderate eccentricities is identified. 
A dynamical analysis of the evolutions is realized in Section \ref{sec:4}, guided by the planar and spatial families of periodic orbits in the dynamical vicinity of the systems. Finally, our results are summarized in Section~\ref{sec:5}.

\section{Final architectures of the IRTP systems} \label{sec:2}


\cite{Sotiriadis2017} have investigated the influence of the eccentricity and inclination damping due to planet-disc interactions on the final configurations of planetary systems, generalizing previous studies on the combined action of the gas disc and planet-planet scattering during the disc phase. {They have performed 11 000 numerical experiments of three giant planets initially on quasi-circular and quasi-coplanar orbits, in the late stage of the gas disc, exploring different initial configurations, planetary mass ratios and disc masses. Following \cite{Libert2011a} and \cite{Teyssandier2014}, disc-induced migration is applied to the outer planet only, since this approach favours convergent migration. Their $n$-body simulations adopted the damping formulae for eccentricity and inclination deduced from the hydrodynamical simulations of \cite{Bitsch2013}. The disc mass was decreased exponentially during the evolution. }

{Although the majority of the multiple systems formed in \linebreak \cite{Sotiriadis2017} are quasi-coplanar, $\sim 5\%$ of them end up with high mutual inclinations ($>10^\circ$). By carefully studying the dynamical evolution of the non-coplanar systems, they have found that $\sim 30\%$ of them result from two- or three-body mean-motion resonance captures, the other $\sim 70\%$ being produced by orbital instability and/or planet-planet scattering. More precisely, for $\sim 14 \%$ of the non-coplanar systems, the inclination increase is produced by a capture in three-body resonance, followed by an inclination-type resonance when the eccentricities are high enough. The remaining $\sim 16\%$ of the highly mutually inclined systems have experienced a scattering/merging event and consist of two planets whose semi-major axes ratio is close to a resonant commensurability.} As previously stated, we denote them as IRTP systems.

\begin{figure*}
\centering
  \includegraphics[width=0.48\textwidth]{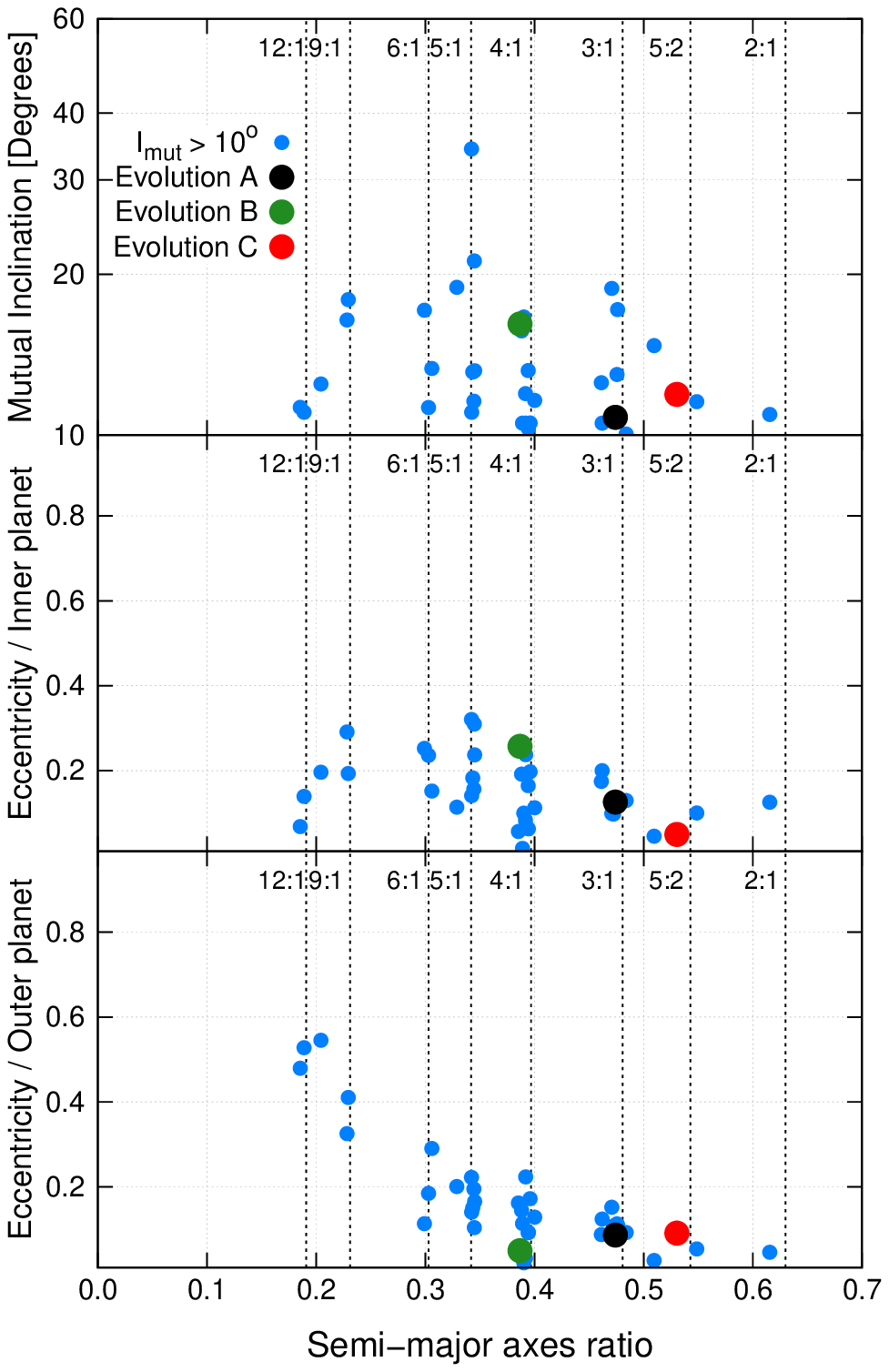} 
  \includegraphics[width=0.49\textwidth]{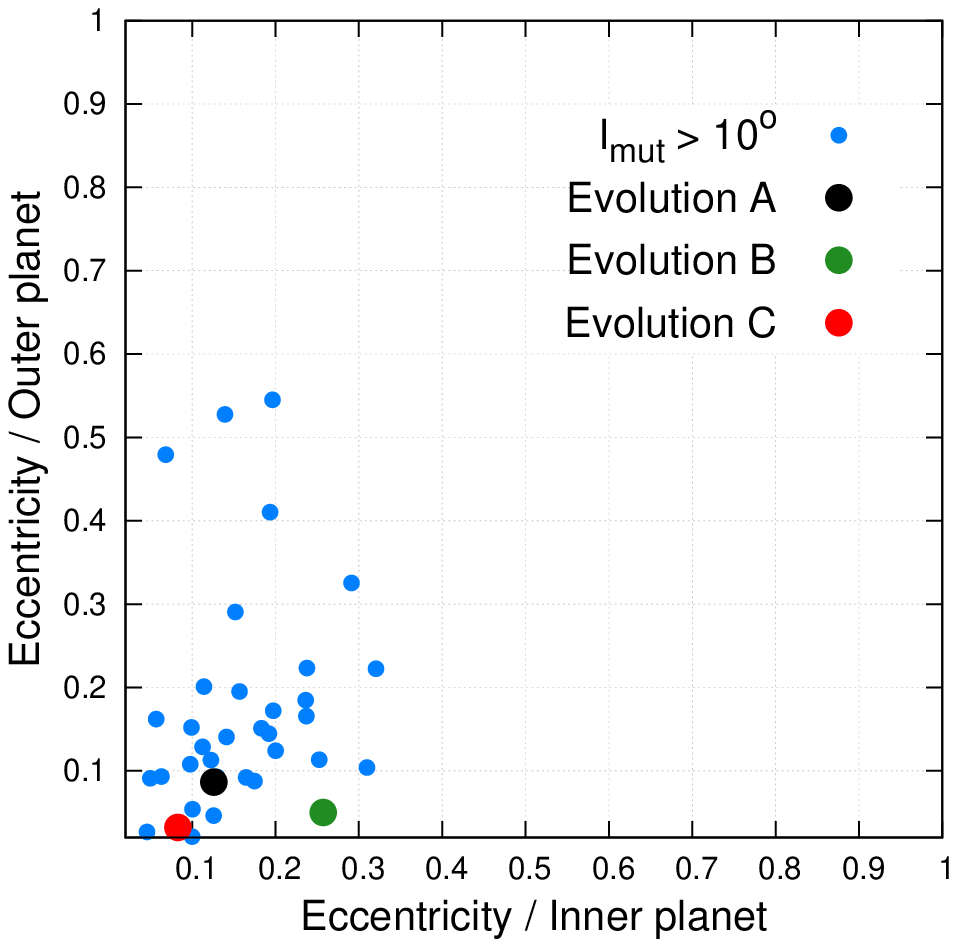}
\caption{Parameters of the IRTP systems of \cite{Sotiriadis2017} at the end of the simulations. Left: From top to bottom, mutual inclination and eccentricities of the inner and outer planets as a function of the semi-major axes ratio. The resonant commensurabilities are indicated with dashed lines. Right: Eccentricities of the inner and outer planets. {The three evolutions analysed in Sections \ref{sec:3} and \ref{sec:4} are also shown.}}
\label{fig:1}       
\end{figure*}

In their simulations, \cite{Sotiriadis2017} have reported $38$ IRTP systems. Their final orbital parameters are shown in Fig. \ref{fig:1}. The proximity of the systems to resonant commensurabilities is obvious (left panel). Interestingly, the systems are mainly gathered around high-order commensurabilities. Concerning the mutual inclination of the systems, the values range from $10^\circ$ to $20^\circ$, except one of them being around $35^\circ$. More puzzling is the limited extent of the eccentricity values (right panel). Most of the systems have their inner and outer planetary eccentricities simultaneously lower than $0.3$. 

{At first glance, these eccentricity values seem to be inconsistent with an inclination-type resonance. Let us recall that in the $n$-body simulations of \cite{Libert2009}, captures in high order resonances leading to inclination excitation were only observed when the eccentricity of one planet exceeds $\sim 0.4$. For elliptic orbits in the 2/1 and 3/1 MMRs, \cite{Voyatzis2014} have confirmed this observation by showing that inclination-type resonance is associated with the existence of vertical critical orbits along the planar {\it elliptic} families of resonant periodic orbits, where families of spatial periodic orbits bifurcate. Recently, regarding the {\it circular} family of periodic orbits, capture in MMR and subsequent inclination-type resonance have been analytically shown for high order MMRs (like 5/2) by \cite{Antoniadou2017}. When following the spatial family emanating from the vertical critical orbit of the circular family, significant increase of the eccentricities and inclinations can be observed. However, this process strongly depends on the value considered for the eccentricity damping.}

Only four systems of Fig. \ref{fig:1} meet the requirement observed in \linebreak \cite{Libert2009}, with the eccentricity of the outer planet being in the range [$0.4$, $0.55$]. These systems are close to very high order commensurabilities (i.e., $9/1$, $11/1$ and $12/1$). {Let us note that the outer planet in the four systems is very massive. The interactions with the gas disc tend to damp a planet in a circular orbit in the case of a low-mass planet and in an orbit whose eccentricity increases over time in the case of a high-mass planet ($> 4 - 5 M_{\rm Jup}$) and a sufficiently massive gas disc, as observed here (see \cite{Bitsch2013}).}

Although the small eccentricities reported in Fig. \ref{fig:1} might be due to the strong damping on eccentricity during the disc phase, how these small to moderate values in eccentricities can coexist with high mutual inclinations deserves careful consideration. In the next section, we present three evolutions of IRTP systems and identify the dynamical mechanism that comes into play to produce the high mutual inclinations.  


\begin{figure}
\centering
  \includegraphics[width=0.75\textwidth]{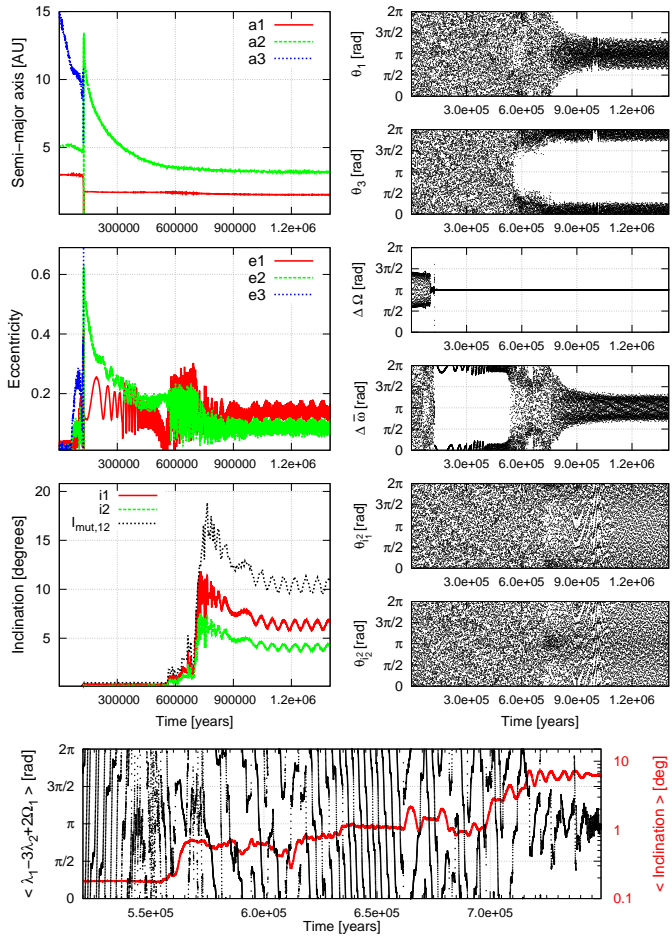}
\caption{Illustration of evolution A. The planetary masses are $m_1=3.49$, $m_2=3.74$, and $m_3=1.42\,M_{\rm Jup}$. The inclination excitation occurs when the planetary system is in the 3/1 MMR and is produced by an inclination-type resonance at low eccentricities. The system is still in MMR at the end of the simulation.}
\label{fig:mecha1}       
\end{figure}
%

\section{Inclination excitation at small to moderate eccentricities} \label{sec:3}

In this section, we describe the formation of three IRTP systems, focusing on their final orbital parameters and the behaviour of their resonant angles. {The orbital elements are computed relative to the invariant plane of the system.}


\paragraph{Evolution A}  

In Fig.~\ref{fig:mecha1}, we present a first typical evolution of a planetary system that shows a sudden growth of the inclinations. During the migration of the outermost planet, the system is destabilized, and the latter planet is rapidly ejected. This scattering event is accompanied by an excitation of the eccentricities of the two remaining planets. However, this excitation is rapidly damped by the gas disc. The system is then captured in 3/1 MMR at around $5.5\times 10^5$ yr, which is characterized by the libration of the resonant angle $\theta_3=\lambda_1-3\lambda_2+\varpi_1+\varpi_2$ around $0^\circ$ (second right panel). {A slightly chaotic evolution follows until the capture in libration of the second resonant angle $\theta_1 = \lambda_1 - 3 \lambda_2 + 2 \varpi_1$ around $180^{\circ}$ (top right panel) and a sudden excitation of the inclinations (up to $\sim 18^\circ$ for the mutual inclination). 

Interestingly, the inclination increase occurs when the two planets have low eccentricities, and no clear capture in an inclination-type resonance can be observed. Nonetheless, from the bottom right panels of Fig.~\ref{fig:mecha1}, we see that the evolutions of the inclination-type resonant angles $\theta_{i_1^2}=\lambda_1-3\lambda_2+2\Omega_1$ and $\theta_{i_2^2}=\lambda_1-3\lambda_2+2\Omega_2$ are slightly perturbated. A further analysis is given in the bottom panel, for the timescale [$5\times 10^5$, $7.5\times 10^5$] yr. To identify the long-period trend when the short-period oscillations are large in amplitude, we have plotted the moving average every $3\times 10^3$ yr of the inclination $i_2$ (red curve, in logarithmic scale) and the inclination-type resonant angle $\theta_{i_2^2}$ (black curve). A clear correlation is observed between the two curves. Also, by removing the fast frequencies, the libration of $\theta_{i_2^2}$ is now clearly visible around $7\times10^5$ yr, when the two resonant angles finally librate simultaneously, indicating the proximity with a vertical critical orbit and a spatial family of periodic orbits, as will be shown in the next section. The 3D configuration is maintained in mean-motion resonance until the end of the simulation, with a mutual inclination of $\sim 10^\circ$.}

\begin{figure}
\centering
  \includegraphics[width=0.75\textwidth]{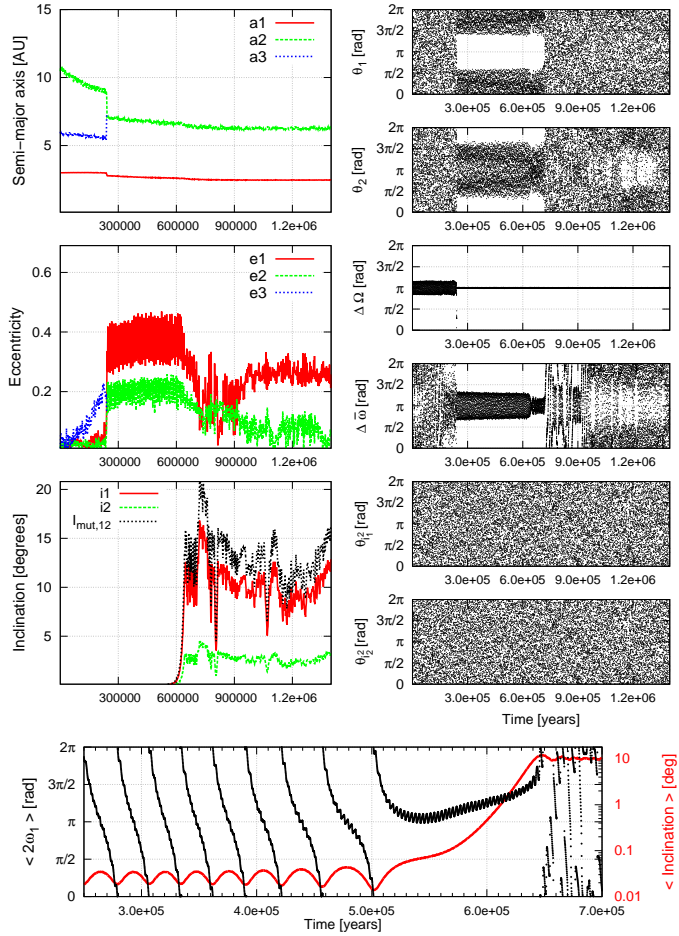}
\caption{Illustration of evolution B. The planetary masses are $m_1=4.14$, $m_2=2.27$, and $m_3=9.51\,M_{\rm Jup}$. The inclination excitation occurs when the planetary system is in the 4/1 MMR and is associated with a libration of $2\omega_1$ (Lidov-Kozai resonance inside the mean-motion resonance). At high mutual inclination, the system is no longer in MMR.}
\label{fig:mecha2}       
\end{figure}

\begin{figure}
\centering
  \includegraphics[width=0.75\textwidth]{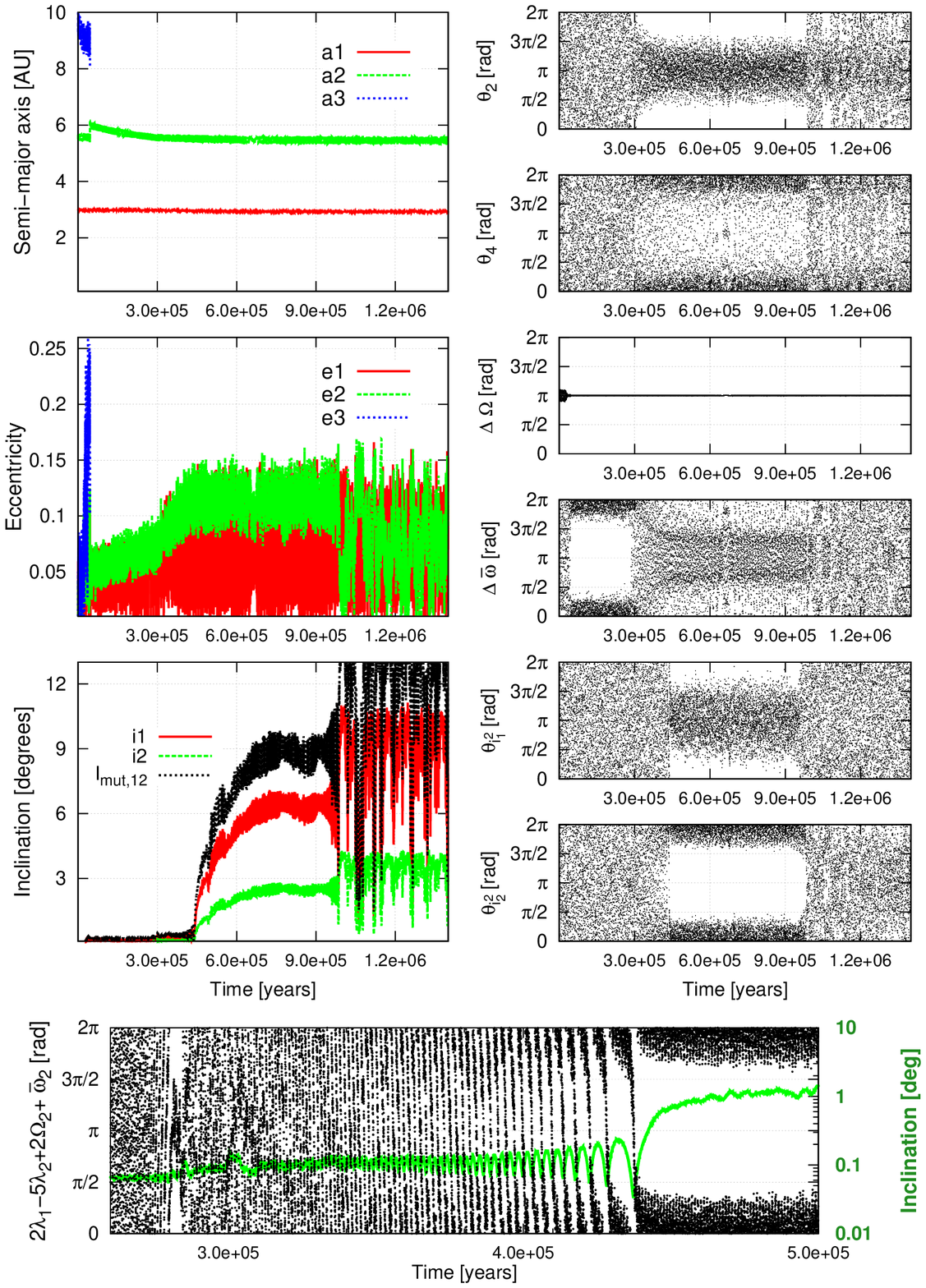}
\caption{Illustration of evolution C. The planetary masses are $m_1=5.63$, $m_2=8.91$, and $m_3=1.66\,M_{\rm Jup}$. The inclination excitation occurs when the planetary system is in the 5/2 MMR and enters an inclination-type resonance at low eccentricities. At high mutual inclination, the system is no longer in MMR.}
\label{fig:mecha3}       
\end{figure}

\paragraph{Evolution B}  

A second evolution is displayed in Fig.$~$\ref{fig:mecha2}. After the ejection of the middle planet, the two-planet system is rapidly captured in 4/1 MMR (at $\sim\!3\times10^5$ years). Since the two resonant angles $\theta_1 = \lambda_1 - 4 \lambda_3 + 3\varpi_1$ and $\theta_2 = \lambda_1 - 4 \lambda_3 + 3\varpi_3$ librate around $0^\circ$ and $180^\circ$, respectively, the planets are in apsidal anti-alignment ($\Delta \varpi$ oscillates around $180^\circ$). Subsequently significant increase in the eccentricities is observed (up to $0.4$ for the eccentricity of the inner planet). At $\sim\!6\times10^5$ yr, the inclinations increase rapidly and the mutual inclination reaches $\sim 20^\circ$. However, the system shows no libration of the inclination-type resonant angles $\theta_{i_1^2}=\lambda_1-4\lambda_3+2\Omega_1+\varpi_1$ and $\theta_{i_2^2}=\lambda_1-4\lambda_3+2\Omega_3+\varpi_3$ (bottom panels, right column). 

{To understand better the inclination increase, we display, in the bottom panel of Fig.$~$\ref{fig:mecha2}, the moving average every $3\times 10^3$ yr of the angle $2 \omega_1$ (black curve), as well as the evolution of the inclination of the inner planet $i_1$ (red curve, in logarithmic scale), during the period of inclination growth. We see that the inclination increases in correlation with the secular libration of $2 \omega_1$. The system seems to follow an invariant curve around the separatrix associated with the Lidov-Kozai dynamics (\cite{Lidov1962,Kozai1962}). Let us remind that the angle $2\omega_1$ is related to the inclination-type resonant angle $\theta_{i_1^2}$, since $\theta_{i_1^2}=\theta_1-2\omega_1$. In the planetary case, an inclination-type resonance can be seen as a Lidov-Kozai resonance embedded in a mean-motion resonance. Thus, the mechanism producing the inclination increase in evolution B is the inclination-type resonance, although the libration of the inclination-type resonant angles shown in Fig. $~$\ref{fig:mecha2} is hidden by the short-period oscillations. Let us note that at high mutual inclination, the system is no longer in MMR.}


\paragraph{Evolution C}

Finally, we show in Fig.$~$\ref{fig:mecha3} an evolution of a planetary system that stays in an inclination-type resonance for a long period of time. During the migration of the outermost planet, the two outer planets merge. At $\sim\!3\times10^5$ yr, the system enters the 5/2 MMR. The resonant angles $\theta_2=2\lambda_1-5\lambda_2+3\varpi_2$ and $\theta_4=2\lambda_1-5\lambda_2+\varpi_1+2\varpi_2$ librate (right column, top panels). The significant inclination increase is again associated here with an inclination-type resonance, as can be deduced from the libration of the angles $\theta_{i_1^2}=2\lambda_1-5\lambda_2+2\Omega_1+\varpi_1$ and $\theta_{i_2^2}=2\lambda_1-5\lambda_2+2\Omega_2+\varpi_2$ (right column, bottom panels). {The correlation between the inclination $i_2$ and the angle $\theta_{i_2^2}$ is shown in the bottom panel of Fig.$~$\ref{fig:mecha3}. The system then evolves along a spatial family of unstable periodic orbits and is finally no longer in MMR, as will be shown hereinafter.}



In the next section, we will explain how the different resonant behaviours highlighted here are linked with families of resonant periodic orbits.   

\section{Influence of the resonant periodic orbits} \label{sec:4}

The families of stable periodic orbits constitute the backbone of stability domains, where the long-term stability is guaranteed. Continuation and existence of periodic orbits of the three-body problem were studied by \linebreak \cite{Hadjidemetriou1975} many years ago, and this work has later found a new field of application in the extrasolar systems (e.g., \cite{Hadjidemetriou2002}, \cite{Beauge2003}, \cite{Antoniadou2016}). Several works have shown that stable periodic orbits can drive the migration process of coplanar planets (e.g., \linebreak \cite{Lee2002}, \cite{FM2003}, \cite{Hadjidemetriou2002}). \linebreak \cite{Voyatzis2014} have studied the spatial case, showing that planetary systems in inclination-type resonance during the disc-induced migration follow families of spatial periodic orbits. Recently,  \cite{Antoniadou2017} have shown the existence of vertical critical orbits for the circular family, as previously discussed.     

Here we aim to analyse the resonant evolutions of the three IRTP systems shown in the previous section. For practical details on the computation of the families of periodic orbits, we refer to \cite{Antoniadou2011} (planar three-body problem) and \cite{Antoniadou2013} (spatial general three-body problem).

\begin{figure}
\centering
  \includegraphics[width=0.48\textwidth]{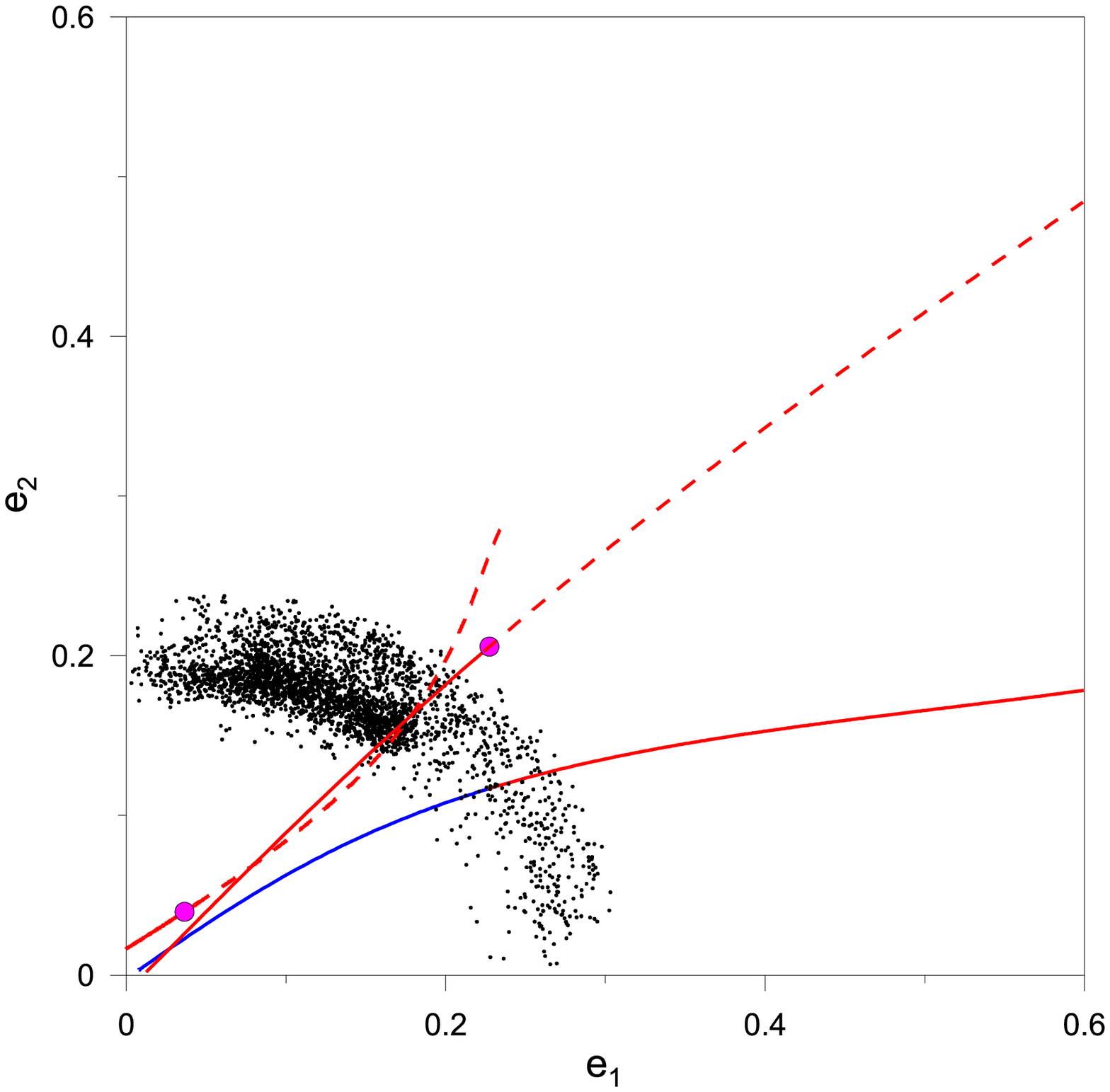}\quad
  \includegraphics[width=0.48\textwidth]{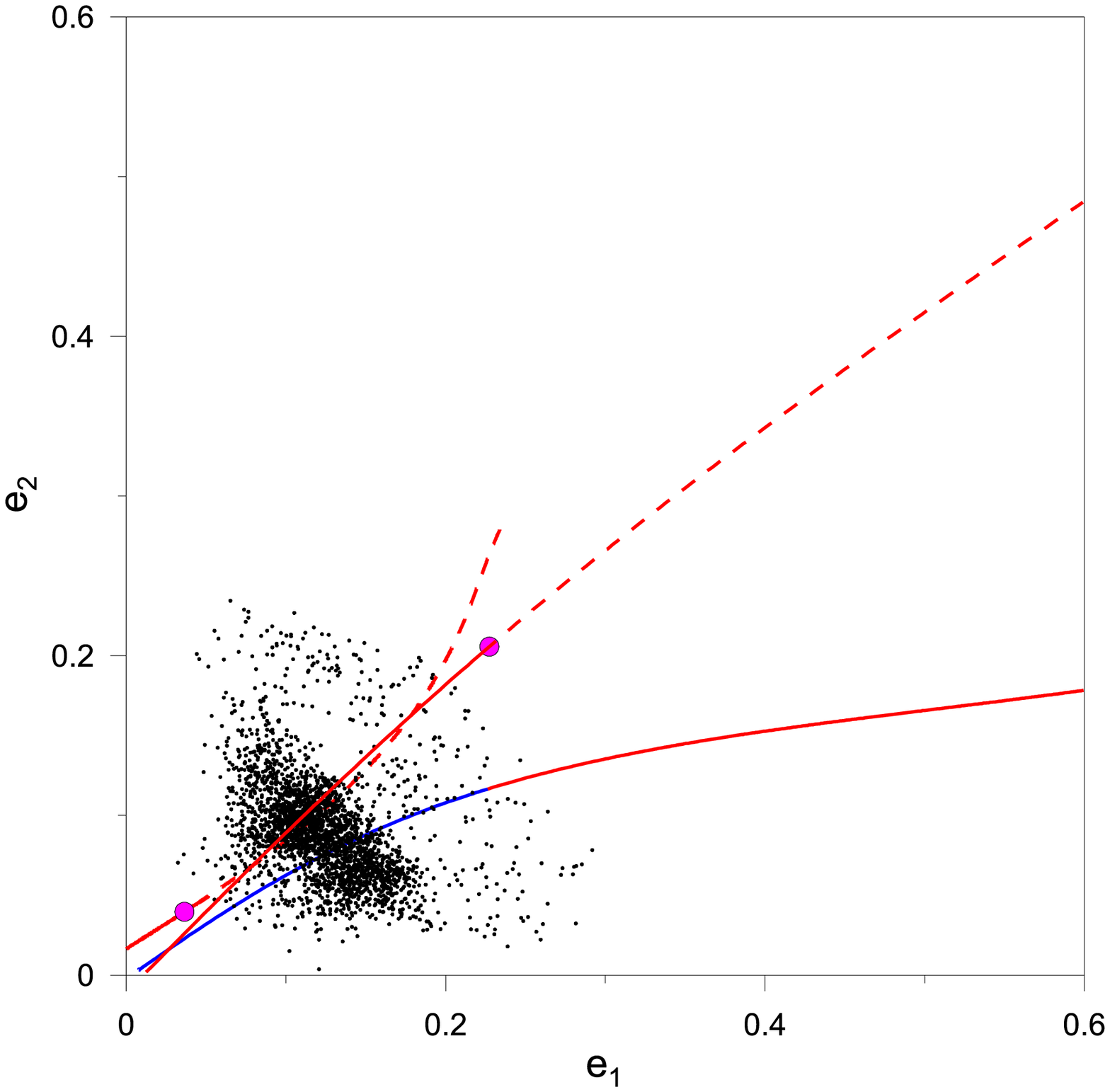}
\caption{Planar families of symmetric periodic orbits in 3/1 MMR related to evolution A (in black dots), on the projection plane $(e_1,e_2)$, for the three configurations $(\theta_3,\theta_1)=(0^\circ,180^\circ)$ (bottom curve), $(\theta_3,\theta_1)=(0^\circ,0^\circ)$ (middle curve) and $(\theta_3,\theta_1)=(180^\circ,0^\circ)$ (top curve). The horizontal stability (instability) is denoted by blue (red) solid lines. Dashed lines correspond to vertically unstable periodic orbits. Left: The evolution of the system in the time period [$4\times10^5$,$7\times10^5$] yr. Right: The evolution of the system in the time period [$7\times10^5$,$10^6$] yr.}
\label{fig:periodic1}       
\end{figure}

\paragraph{Evolution A}
The planar elliptic families of symmetric periodic orbits in 3/1 MMR and their bifurcations to spatial families have been investigated in \linebreak \cite{Antoniadou2014} (see their Figure 6). We display in Fig.$~$\ref{fig:periodic1} the families with resonant configurations $(\theta_3,\theta_1)=(0^\circ,180^\circ)$ (bottom curve), $(\theta_3,\theta_1)=(0^\circ,0^\circ)$ (middle curve) and $(\theta_3,\theta_1)=(180^\circ,0^\circ)$ (top curve), for the mass ratio $m_1/m_2=0.93$. Blue lines represent (horizontally) stable families, while the red ones (horizontally) unstable families. Coloured dots indicate the vertical critical orbits where families of spatial periodic orbits bifurcate\footnote{Colours refer to the symmetry of the spatial periodic orbits they generate (see \cite{Antoniadou2014} for more details).}. {The evolution of the planetary eccentricities of system A is also shown with black dots before the inclination increase (left panel) and at the moment of the inclination increase (right panel). 

At the beginning of the evolution, the system is in apsidal alignment around $0^\circ$, until the capture in the 3/1 MMR at $5.5\times10^5$ yr associated with the libration of the resonant angle $\theta_3=\lambda_1-3\lambda_2+\varpi_1+\varpi_2$ around $0^\circ$. Since the angle $\theta_1=\lambda_1-3\lambda_2+\varpi_1$ rotates, the system alternatively crosses the three planar families of periodic orbits, which are all horizontally unstable at these eccentricities. The family with resonant configuration $(\theta_3,\theta_1)=(180^\circ,0^\circ)$ possesses a vertical critical orbit at low eccentricities. The latter was not reported in \cite{Voyatzis2014} because they studied the differential migration of two planets starting from quasi-circular and quasi-coplanar orbits which followed the stable elliptic families. In the right panel of Fig.$~$\ref{fig:periodic1}, we plot the evolution of the system when its mutual inclination increases due to the proximity to this vertical critical orbit. Note that the spatial family close to which the non-coplanar system evolves could not be computed here, since the planar family of periodic orbits is very unstable to be continued at space.} 

\begin{figure}
\centering
  \includegraphics[width=0.48\textwidth]{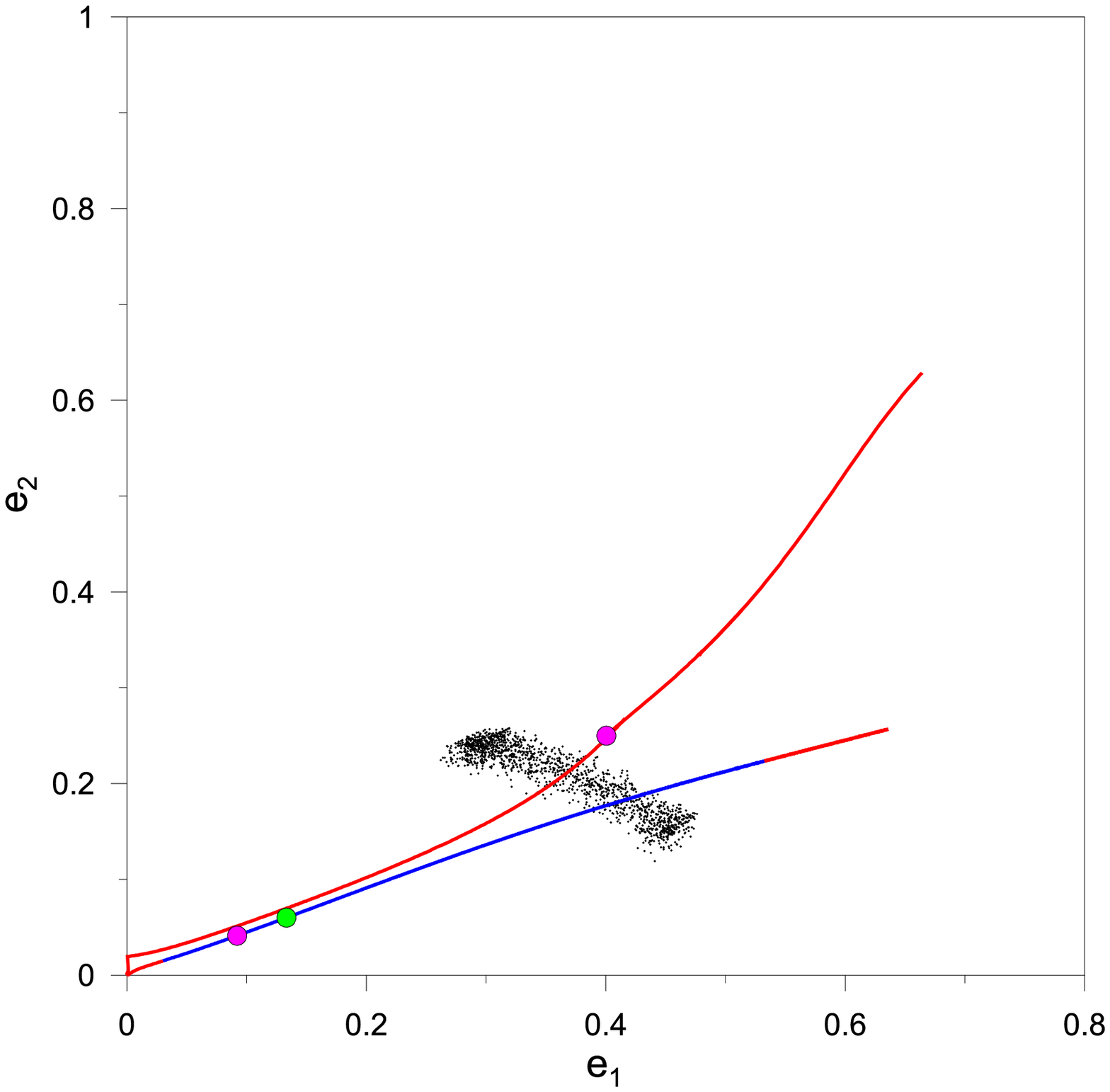}\quad
  \includegraphics[width=0.48\textwidth]{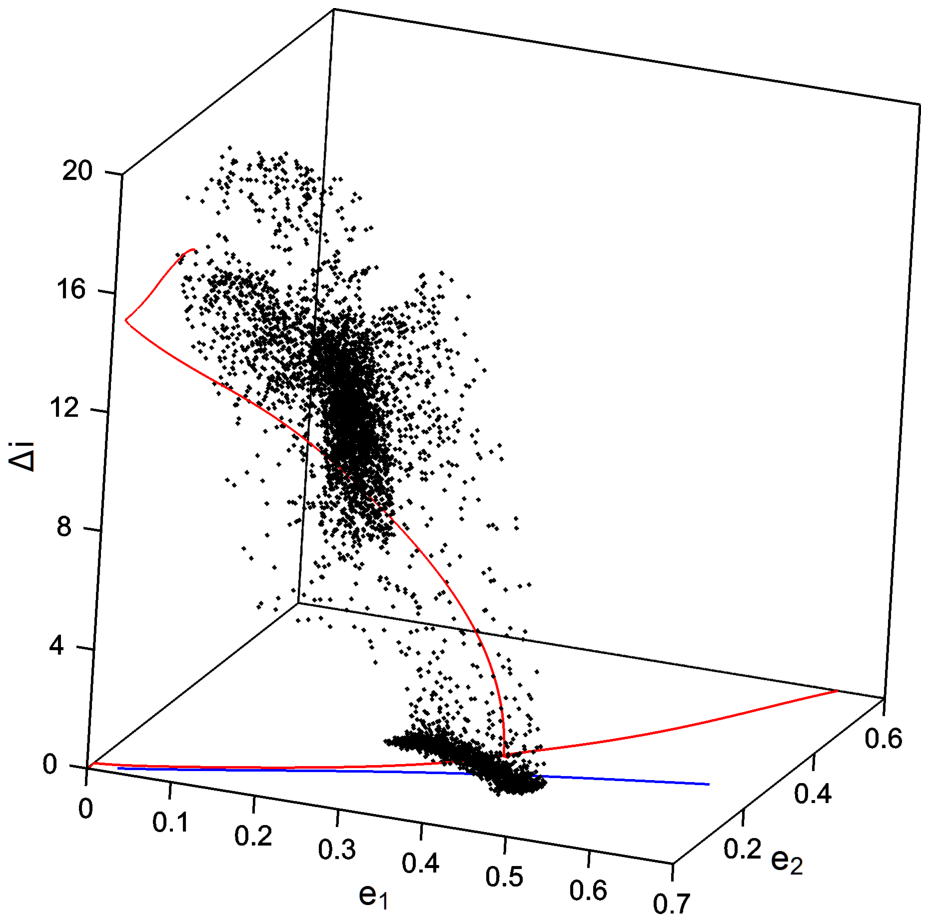}
\caption{Left: Planar families of symmetric periodic orbits in 4/1 MMR related to evolution~B (in black dots, the first $6\times 10^5$ yr only), on the projection plane $(e_1,e_2)$, for the two configurations $(\theta_1,\theta_2)=(0^\circ,180^\circ)$ (bottom curve) and $(\theta_1,\theta_2)=(180^\circ,0^\circ)$ (top curve). Right: Spatial family of unstable periodic orbits emanating from the vertical critical orbit of the  family $(\theta_1,\theta_2)=(180^\circ,0^\circ)$, on the projection plane $(e_1,\Delta i)$.}
\label{fig:periodic2}       
\end{figure}

\paragraph{Evolution B}

In Fig. \ref{fig:periodic2} (left panel), we show the planar families of symmetric periodic orbits in 4/1 MMR  for the two configurations $(\theta_1,\theta_2)=(0^\circ,180^\circ)$ (bottom curve) and $(\theta_1,\theta_2)=(180^\circ,0^\circ)$ (top curve), when $m_1/m_3=0.44$ (see Figure 7 of \cite{Antoniadou2014}). The black dots represent the evolution of the planetary eccentricities before the inclination increase at $\sim~6 \times 10^5$ yr. {For the eccentricity values of evolution B, during the first $6\times 10^5$ yr, the planar family associated with $(\theta_1,\theta_2)=(0^\circ,180^\circ)$ is both horizontally and vertically stable, unlike the planar family associated with $(\theta_1,\theta_2)=(180^\circ,0^\circ)$, which is horizontally unstable and vertically stable. Let us note the existence of a vertical critical orbit on the latter family at eccentricities close to the ones of evolution~B. While evolving along the two families, the system is attracted by the spatial family of unstable periodic orbits emanating from this vertical critical orbit (right panel of Fig. \ref{fig:periodic2}). When reaching high mutual inclination, the system gets out of the MMR and shows a chaotic evolution.}

\begin{figure}
\centering
  \includegraphics[width=0.48\textwidth]{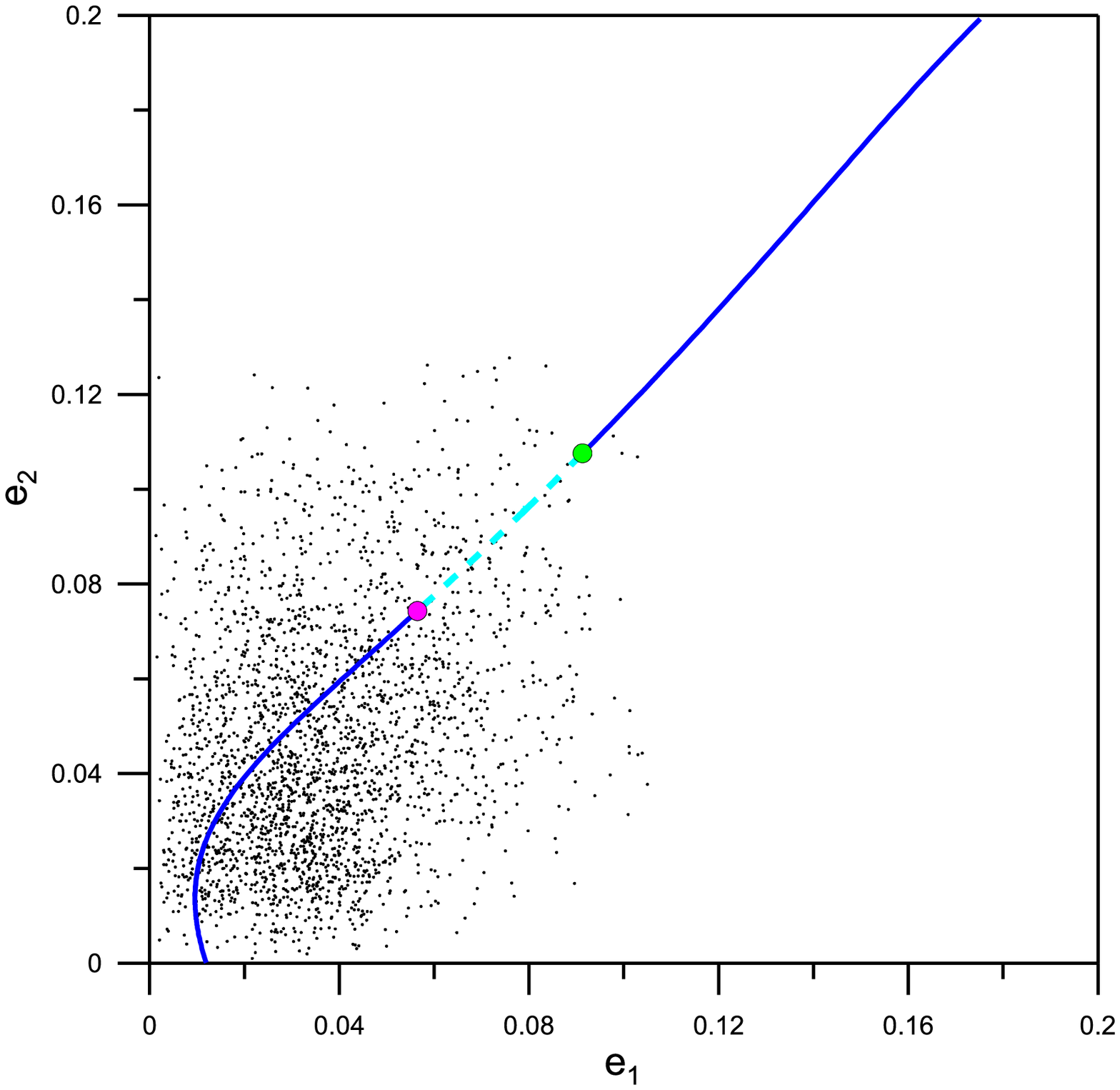}  \quad
  \includegraphics[width=0.48\textwidth]{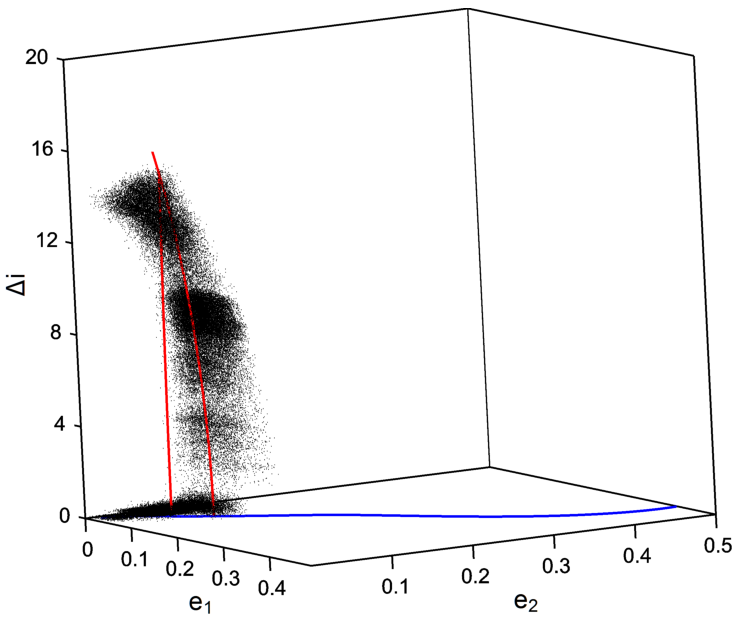}
\caption{Left: Planar family of symmetric periodic orbits in 5/2 MMR related to evolution~C, on the projection plane $(e_1,e_2)$. The system initially evolves along the planar family of periodic orbits, until reaching a vertical critical orbit (green dot). Right: Spatial families of periodic orbits emanating from the two vertical critical orbits at small eccentricities, on the projection plane $(e_1,\Delta i)$. The system enters an inclination-type resonance and follows the (unstable) spatial family.}
\label{fig:periodic3}       
\end{figure}

\paragraph{Evolution C}

The planar family of symmetric periodic orbits in 5/2 MMR for $(\theta_1,\theta_2)=(0^\circ,180^\circ)$ (mass ratio of $0.53$) is displayed in the left panel of Fig.$~$\ref{fig:periodic3} (see also Figure 5 of \cite{Antoniadou2014}). {The system migrates along the planar family of periodic orbits, which is horizontally stable, but vertically unstable between the two vertical critical orbits (dashed line) at small eccentricities. It first reaches the vertical critical orbit at $e_1=0.06$ and $e_2=0.07$, and acquires a small inclination increase at $\sim 3 \times 10^5$ yr. Then, following the planar family, it meets the other vertical critical orbit at $e_1=0.09$ and $e_2=0.11$, and enters an inclination-type resonance along the unstable spatial family emanating from the latter vertical critical orbit (right panel of Fig. \ref{fig:periodic3}). The system is destabilized when the mutual inclination approaches $\sim 10^{\circ}$ and eventually gets out of the 5/2 MMR.}

\section{Conclusions} \label{sec:5}

In the present work, we have shown that the dynamics of planetary systems after an ejection or a merging during the disc-induced migration is extremely rich and complex. {By carefully analysing several dynamical evolutions, we have pointed out that inclination excitation can be produced at small to moderate eccentricities by a (temporary) inclination-type resonance. This resonance is associated with a vertical critical orbit along a planar elliptic family of periodic orbits. Our study shows that no minimum values of the eccentricities are required for the establishment of an inclination-type resonance in the elliptic case\footnote{A similar observation was previously made by \cite{Antoniadou2017} for the planar circular family (i.e. capture in MMR and inclination-type resonance can occur when both eccentricities are close to $0$).}.} This mechanism operates in all the $38$ IRTP systems of \cite{Sotiriadis2017}.

The influence of periodic orbits on the final parameters of the system is crucial, as highlighted by the three  evolutions considered in this work. {They draw preferred paths in eccentricities and inclinations for the evolution of the systems. The proximity to spatial periodic orbits contributes to maintaining the mutual inclination of the systems over long periods of time.}

Let us also note that the joint action of an ejection/merging and an \linebreak inclination-type resonance, as described hereabove, can even drive a planetary system to stable regions that could not have been reached by migration from quasi-circular and quasi-coplanar orbits.$~$It is the case for evolution B for instance, which initially evolves around a planar family of periodic orbits that is unstable at small eccentricities.$~$The influence of the resonant mechanisms on the past history of planetary systems found in the vicinity of a resonant commensurability should not be underestimated, but be deeply analysed in formation studies aiming to explain the parameters of the detected extrasolar planets.

\begin{acknowledgements}
The authors would like to thank K. Tsiganis and A. Morbidelli for useful discussion. This work was supported by the Fonds de la Recherche Scientifique-FNRS under Grant No. T.0029.13 (“ExtraOrDynHa” research project). Computational resources have been provided by the Consortium des \'Equipements de Calcul Intensif (C\'ECI), funded by the Fonds de la Recherche Scientifique de Belgique (F.R.S.-FNRS) under Grant No. 2.5020.11.   
\end{acknowledgements}



\end{document}